\documentclass[useAMS,  usenatbib]{mn2e}
\usepackage{amsmath}
\usepackage{cases}
\usepackage{txfonts}
\usepackage{tipa}
\usepackage{amsfonts}
\usepackage{amssymb}
\usepackage{graphicx}
\usepackage{multirow}

\title[Photometric Metallicity Calibration]{Photometric Metallicity Calibration with SDSS and SCUSS and its Application to distant stars in the South Galactic Cap}

\author[Gu et al.]{Jiayin Gu$^{1}$, Cuihua Du$^{1}$ \thanks{Email: ducuihua@ucas.ac.cn (Du)}, Yunpeng Jia$^{2}$, Xiyan Peng$^{2}$, Zhenyu Wu$^{2}$,  Yingjie Jing$^{2}$,
\and Jun Ma$^{2}$, Xu Zhou$^{2}$,  Xiaohui Fan$^{3}$, Zhou
Fan$^{2}$,Yipeng Jing$^{4}$,Zhaoji Jiang$^{2}$, Michael
Lesser$^{3}$,
\and  Jundan Nie$^{2}$, Shiyin Shen$^{5}$, Jiali Wang$^{2}$, Hu Zou$^{2}$, Tianmeng Zhang$^{2}$, Zhimin Zhou$^{2}$,  \\
$^{1}$School of physics sciences, University of the Chinese Academy of Sciences, Beijing 100049, P. R. China \\
$^{2}$Key Laboratory of Optical Astronomy, National Astronomical Observatories, Chinese Academy of Sciences, Beijing, 100012, China \\
$^{3}$Department of Astronomy and Steward Observatory, University of Arizona, Tucson, Arizona, USA \\
$^{4}$Department of Physics and Astronomy, Shanghai Jiao Tong University, Shanghai 200240 \\
$^{5}$Shanghai Astronomical Observatory, Chinese Academy of Sciences, Shanghai 200030 \\
}

\begin{document}

\maketitle

\begin{abstract}
\par Based on SDSS g, r and SCUSS (South Galactic Cap of u-band Sky Survey) $u$ photometry,
we develop a photometric calibration for estimating the stellar
metallicity from $u-g$ and $g-r$ colors by using the SDSS spectra of
32,542 F- and G-type main sequence stars, which cover almost $3700$
deg$^{2}$ in the south Galactic cap. The rms scatter of the photometric
metallicity residuals relative to spectrum-based metallicity is $0.14$ dex when $g-r<0.4$,
and $0.16$ dex when $g-r>0.4$. Due to the deeper and more
accurate magnitude of SCUSS $u$ band, the estimate can be used up to
the faint magnitude of $g=21$. This application range
of photometric metallicity calibration is wide enough so that it can
be used to study metallicity distribution of distant stars. In this
study, we select the Sagittarius (Sgr) stream and its neighboring
field halo stars in south Galactic cap to study their metallicity
distribution. We find that the Sgr stream at the cylindrical
Galactocentric coordinate of $R\sim 19$ kpc, $\left| z\right| \sim
14$ kpc exhibits a relative rich metallicity distribution,
and the neighboring field halo stars in our studied fields can be
modeled by two-Gaussian model, with peaks respectively at
[Fe/H]$=-1.9$ and [Fe/H]$=-1.5$.
\end{abstract}


\section{Introduction}
\par In the standard hierarchical model of galaxy formation, galaxies like
the Milky Way were constructed by repeated aggregation with dwarf galaxies
and this merging process left behind many streams, satellites,
and substructures in the Galactic halo \citep[][]{SearleZinn78}.
The Two Micro All-Sky Survey (2MASS) and
Sloan Digital Sky Survey \citep[SDSS;][]{Abazajian09} provided astronomers
a great opportunity to detect the substructures in the spheroid of the
Milky Way photometrically and spectroscopically \citep[][]{Yanny00,  Newberg02, Majewski03,
Ivezic00, Yanny03, Newberg07, GrillmairDionatos06, Willett09, Belokurov06a, Belokurov06b,
Grillmair06a, Grillmair06b, Belokurov07, Grillmair09}. Among all the substructures,
the most prominent is the tidal stream from the disrupting Sgr dwarf spheroid galaxy
which was discovered by \cite{Ibata94}. Others include the Monoceros Ring \citep[][]{Yanny03},
the Hercules-Aquila Cloud \citep[][]{Belokurov07} and overdensity in the direction of Virgo \citep[][]{Juric08}, and so on.

\par The Sloan Extension for Galactic Understanding and Exploration (SEGUE) Survey
obtained about 240,000 low resolution spectra of faint stars
($14.0<g<20.0$) of a wide variety of spectral types. The limited
number of spectra is far from enough to study the metallicity
distribution of a vast area of the Galaxy. The advantage of using
the photometric metallicity is that the metallicity of many more
stars can be obtained. \cite{Ivezic08} derived a photometric
metallicity relations in $u-g$ vs. $g-r$ plane by using F-and G-type
main sequence stars, and studied the metallicity distribution of the
Galaxy. However, the SDSS $u$ band photometry is limited to
$u\sim22$. Additionally, due to the relatively large error of SDSS
$u$ band magnitude near the faint end, the application of the
photometric metallicity estimates is greatly restricted in the range
of $g<19.5$, an insufficient depth to explore distant substructures.
Based on a calibration of less metallicity-sensitive color indexes
in the $gri$ passbands with well-determined spectroscopic metal
abundances, \cite{An09} determined the metal abundance estimates for
main-sequence stars in the Virgo Overdensity. \cite{Karaali11}
used the UBV photometry to derive the improved metallicity Calibration.
In this study,  we will use a deep photometric survey, South Galactic Cap of
$u$-band Sky Survey (SCUSS), to give a photometric calibration for
estimating the stellar metallicity. The SCUSS $u$ is 1.5 mag deeper
than SDSS $u$. This deep SCUSS $u$-band data can probe a much larger
volume of the halo that enable us to study the MDF of distant
stream.

\par Over the past decades, considerable efforts have also been made to
gain information about the metallicity distribution of
the Sgr system \citep[][]{Smecker02, Bonifacio04, Monaco03, Monaco05, Vivas05, Chou07}.
Many works on metallicity of Sgr stream are based either on high-resolution spectra of
a small amount of stars or on low-resolution spectra of giant stars.
Based on high-resolution spectra, \cite{Bellazzini08} selected 321 RGB stars
in the Sgr nucleus and give the mean metallicity of $-0.45$ dex from the infrared CaII triplet.
\cite{Yanny09} traced the Sgr stream with 33 red K/M giant stars from SDSS low-resolution spectra
and found an mean metallicity in the range of $-0.8\pm 0.2$. Using CFHTLS data,
\cite{Sesar11} detected the Sgr stream in $l=173^\circ$ and $b=-62^\circ$
as an overdensity of [Fe/H]$\sim -1.5$ at $R_{gal}\sim 32$ kpc. \cite{Shi12} selected 556 red
horizontal branch stars along the Sgr streams from SDSS DR7 spectroscopic data, and found that
the Sgr stars have two peaks in the metallicity distribution while the Galactic stars have a more prominent
metal-poor peak. Up to now, the Sgr MDF remain controversial due to different studies. In this study,
we attempt to estimate the photometric MDF of the Sgr stream in south Galactic cap,
which is as an application of the SCUSS photometric metallicity calibration.
Additionally, we also evaluate the MDF of the field halo stars which are in the vicinity of the Sgr stream.

\par The outlines of this paper are as follows.
We take a brief overview of the SDSS and SCUSS in Section 2,
and give the SCUSS photometric metallicity calibration in Section 3.
Section 4 presents the overdensities in the south Galactic cap.
In Section 5, we discuss the photometric metallicity distribution of
the Sgr stream in certain region and its neighboring field. A summary is given in Section 6.

\section{An overview of the SDSS and SCUSS}

\par The SDSS is a digital multi-filter imaging and spectroscopic redshift survey using a dedicated 2.5 m telescope \citep{Gunn06}.
It covers more than one-quarter of the celestial sphere in the north
Galactic cap, as well as a small ($\sim 300$ deg$^{2}$) but much
deeper survey in the south Galactic hemisphere. It uses five bands
($u$, $g$, $r$, $i$, and $z$) to simultaneously measure brightness
of stars with the effective wavelength 3562, 4686, 6165, 7481, and
8931 \AA~ respectively. Its magnitude limits (95\% completeness for
point source) are 22.0, 22.2, 22.2, 21.3, and 20.5 for $u$, $g$,
$r$, $i$, and $z$ respectively \citep[][]{Abazajian04}. The relative
photometric calibration accuracy for $u$, $g$, $r$, $i$, and $z$ are
2\%, 1\%, 1\%, 1\% and 1\% respectively \citep[][]{Padmanabhan08}.
The other technical details about SDSS can be found on the SDSS
website \emph{http://www.sdss3.org/}, which also provide interface
for the public data access.

\begin{figure}
\includegraphics[width=1.0\hsize]{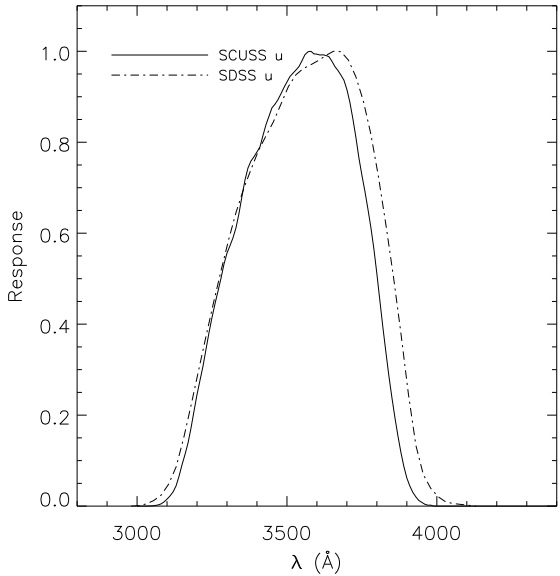}
\caption{Response curves of both the SCUSS $u$ and the SDSS $u$ filters. Both curves include the atmospheric extinction at the airmass of 1.3 and normalized to their maxima. The SCUSS $u$ band filter has about 24 \AA ~blue shift and is slightly narrower than SDSS $u$ band filter.}
\label{Response}
\end{figure}

The South Galactic Cap u-band Sky Survey (SCUSS) is an international cooperative project,
which is undertaken by National Astronomical Observatories, Chinese Academy of Sciences
and Steward Observatory, University of Arizona, USA. It is an $u$ band (3538 \AA) imaging survey program
with the 90 inch (2.3 m) Bok telescope located on Kitt Peak, near Tucson, AZ.
It will also provide part of the essential input data to the Large Sky Area Multi-Object
Fiber Spectroscopic Telescope (LAMOST) project \citep[][]{Zhao06}.
The $u$ band filter used in SCUSS project is similar to SDSS $u$ band filter but a little bluer. \textbf{Fig.~\ref{Response} displays both SCUSS and SDSS system response functions. The effective wavelength of the SCUSS $u$ band is about 3538 \AA, and the full width half maximum (FWHM) is about 520 \AA.}
By testing the data obtained from previous camera and new 90 Prime camera of BOK telescope,
the limiting magnitude of SCUSS $u$ band may be 1.5 mag deeper than that of SDSS $u$ \citep[][]{Jia14,Peng15}.
The SCUSS project began its observation in 2010 and ended in 2013.
Finally, 3700 square degrees field in the south Galactic cap ($30^\circ<l<210^\circ,~ -80^\circ<b<-20^\circ$) were surveyed.
The SCUSS can be used to study Star formation rate, Galactic interstellar extinction, Galaxy morphology, the Galaxy structure,
Quasi-Stellar Object, Variable star and Cosmology. In Table~\ref{tab_compare},
we list the parameters of SCUSS and SDSS filters. Column (1) represents the ID of SCUSS and SDSS filters.
and columns (2) and (3) represent effective wavelengths and full width at half maximum (FWHM) of six filters,
respectively \citep{Jia14}. The detailed information and data reduction about SCUSS can be found in \cite{Zhou15} and \cite{Zou15a}.

\begin{table}
\begin{center}
\caption{Parameters of SCUSS and SDSS filters. Column (1) represents the ID of SCUSS and SDSS filters, and columns (2) and (3) represent effective wavelengths and full width at half maximum (FWHM) of six filters, respectively.}
\begin{tabular}{p{1.8cm}p{1.8cm}p{1.8cm}}
\hline
\hline
Filter & Wavelength & FWHM \\
       & (\AA)    &(\AA)\\ \hline
 $u$ (SCUSS)  & 3538   &  520      \\ \hline
 $u$ (SDSS)   & 3562   &  575  \\ \hline
 $g$   & 4686   &  1390   \\ \hline
 $r$   & 6165   &  1370  \\ \hline
 $i$   & 7481   &  1530  \\ \hline
 $z$   & 8931   &  950   \\ \hline
\hline
\end{tabular}
\label{tab_compare}
\end{center}
\end{table}

\par The relation between SDSS $u$ and SCUSS $u$ can be expressed by following polynomial transformation:
\begin{align}
u_{\rm SCUSS}=&u_{\rm SDSS}-0.1373(u_{\rm SDSS}-g) \nonumber \\
&+0.0438(u_{\rm SDSS}-g)^{2}+0.1036 \nonumber
\end{align}
where $0.8<u_{\rm SDSS}-g<2.8$. The corresponding error of the
transformed SCUSS $u$ magnitude is estimated by error transfer. The
transformation is derived by point sources with SCUSS and SDSS
photometric errors less than 0.05 mag. It is applied to point
sources, and maximum systematic difference between SCUSS and SDSS is
about 0.036 mag \citep{Zou15b}.

\section{Photometric metallicity calibration}

\par In the SCUSS database, each star has a $u$ band magnitude and its corresponding error,
as well as the position ($ra~\&~dec$). We can identify the same star from the SDSS by matching the position.
Each star also has its unique ID, $u,g,r,i,z$ band magnitude, photometric error and extinction which are derived
from SDSS database (hereafter, when not specified, we use $u$ to only denote the magnitude from SCUSS).
Throughout this paper we use the extinction values from \cite{Schlegel98} which are provided in the SDSS database.

\par The stellar parameters, such as effective temperature, surface gravity,
and metallicity (parameterized as [Fe/H]) are derived with
sufficient accuracy from SDSS stellar spectra. Since the number of
stars from spectroscopic survey  is far smaller than those from
photometric survey, it is unable to get a whole picture of stellar
metallicity distribution from vast area. Considering that the
exhaustion of metals in a stellar atmosphere has detectable effect
on the emergent flux \citep[][]{Schwarzschild55}, in particular in
the blue region where the density of metallicity absorption is
highest, the combination of spectroscopic data and photometric data
can derive the estimates of [Fe/H] \citep[][]{Allende06, Allende08,
Lee08a, Lee08b}. \cite{Ivezic08} derived a metallicity estimator
using third-order polynomials. We adopt the procedure suggested in
\cite{Ivezic08} to derive a metallicity estimator only with the
exception that we use SCUSS $u$ magnitude rather than SDSS $u$. At
first, we get the adopted stellar parameters (estimated through the
SEGUE Stellar Parameter Pipeline [SSPP] \cite[][]{Beers06} of the
stellar spectroscopic survey from sppParams table in SDSS website
(\emph{http://www.sdss3.org/}), and then find the same stars from
SCUSS database by matching the shared ID number of stars. After
excluding the repeated stars surveyed in different plates, we
eventually obtain a database of 74,133 stars with SCUSS $u$
magnitude, SDSS $g,r,i, and~z$ magnitude, as well as the SEGUE
stellar parameters.

\par In order to derive the photometric metallicity estimate, we adopt the same selection criteria used by \cite{Ivezic08}.
We relist these criteria as follows:

\begin{itemize}
\item The interstellar extinction in the $r$ band below 0.3;
\item $14<g<19.5$;
\item $0.2<(g-r)<0.6$;
\item $0.7<(u-g)<2.0$ and $-0.25<(g-r)-0.5(u-g)<0.05$;
\item $-0.2<0.35(g-r)-(r-i)<0.10$;
\item $log (g)>3$.
\end{itemize}

\par Through these selection criteria, we eventually get 32,542 stars to determine a photometric metallicity ([Fe/H]$_{phot}$) estimate.
The median metallicity of the above selected sample stars shows a complex behavior as a function of $u-g$ and $g-r$ colors
which is consistent with the result of \cite{Ivezic08}. To reliably estimate the photometric metal abundance,
it is necessary that these stars are separated by $g-r=0.4$ into two regions in the $u-g$ versus $g-r$ plane to avoid relative large systematic errors.
Then, we adopt third-order polynomials to fit the median metallcity with the color $u-g$ and $g-r$. The resulting estimate is as follows:

when    ~~~~~~ $y<0.4$
\begin{align}
\rm[Fe/H]=&17.68-91.34x+60.55y+110.7x^2 \nonumber \\
&-65.02xy-71.72y^2-41.54x^3 \nonumber \\
&+33.42x^2y-3.734xy^2+58.92y^ 3\nonumber
\end{align}

when    ~~~~~ $y>0.4$
\begin{align}
\rm [Fe/H]=&12.89-6.322x-79.79y+17.06x^2 \nonumber \\
&-38.62xy+199.1y^2-18.85x^3 \nonumber \\
&+102.8x^2y-212.8xy^2+38.17y^3 \nonumber
\end{align}
where $x=u-g$, $y=g-r$.

\par
\begin{figure}
\begin{center}
\includegraphics[width=1.0\hsize]{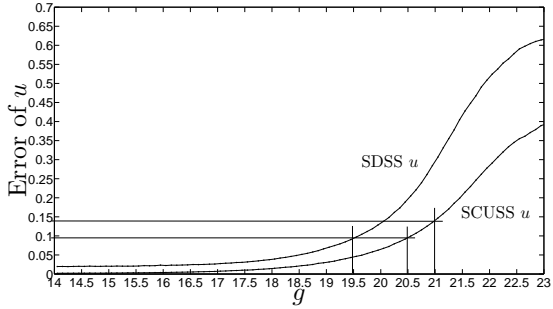}
\caption{The average error of $u$ magnitude of 50,000 stars as a function of $g$ magnitude. The SCUSS $u$ error is smaller than SDSS $u$ error on the whole. From this figure we see that SDSS $u$ error of 0.09 corresponds to $g$ magnitude of 19.5. However the same error of SCUSS $u$ corresponds to deeper $g$ magnitude of 20.5. The $g$ magnitude of 21 corresponds to SCUSS $u$ error of 0.13, which may introduce the rms scatter of photometric metallicity estimate up to $0.445$ dex.}
\label{Error}
\end{center}
\end{figure}

 Using the above equations,  the photometric metallicty [Fe/H]$_{phot}$ is
computed for each star and compared with the spectroscopic
metallicity. The rms scatter of the photometric metallicity
residuals relative to spectrum-based metallicity is $0.14$ dex when
$g-r<0.4$, and $0.16$ dex when $g-r>0.4$. So the photometric
estimates are relatively reliable to provide a good estimation for
the large number of photometric surveyed stars. The random
metallicity error mainly comes from the photometric error of $u$
band magnitude.  As shown in Fig.~\ref{Error}, the average error of
SCUSS $u$ magnitude in most observation field (here, about 50,000
stars as an example) is small than SDSS $u$ on the whole. From this
figure we see that SDSS $u$ error of 0.09 corresponds to $g$
magnitude of 19.5. However, the same error of SCUSS $u$ corresponds
to deeper $g$ magnitude of 20.5. The $g$ magnitude of 21 corresponds
to SCUSS $u$ error of 0.13. Due to its relative deep magnitude and
high accuracy from SCUSS data, we expect wide range of application
of the photometric estimate. We obtain a star sample from the SCUSS
database under the criteria mentioned above except $log (g)>3$, and
compute the rms statistical scatter of [Fe/H]$_{phot}$ introduced by
the $u$ band photometric error. The rms statistical scatter of
metallicity increases with $g$ magnitude from $0.014$ dex for $g<17$
to $0.070$ dex at $g=18.5$, $0.379$ dex at $g=20.5$, $0.445$ dex at
$g=21$ and $0.750$ dex at $g=21.3$. So we limit the application of
photometric estimates in the range of $g<21$, where the maximum rms
scatter of the metallicity is $0.445$ dex. This application of
[Fe/H]$_{phot}$ allows the metallicity to be determined for all
SCUSS stars in these criteria range. So we can derive the
[Fe/H]$_{phot}$ for numerous farther stars.

\par To show how well the present SCUSS photometric metallicity predictions
agree with SDSS photometric metallicity predictions, here as an example,
we randomly select a few main sequence stars, which are bright enough (with $g<19$) to
be well-detected in both SDSS $u$ and SCUSS $u$, to evaluate their metallicity,
respectively from SDSS $u$ and SCUSS $u$ (see Table.~\ref{Metallicity_comparison}).
In Table~\ref{Metallicity_comparison}, Column (1) shows the g magnitude bins,
and the rest columns in each row give the mean values of information about
a bunch of stars which belong to the corresponding bins. The last two columns
give the mean metallicity values estimated respectively from SDSS $u$ and SCUSS $u$.
From the table we can easily find that SCUSS $u$ photometric metallicity estimate is almost
consistent with the SDSS $u$ photometric metallicity estimate.
Nevertheless, despite agreements in both photometric metallicity estimates at bright magnitude,
there still exists mismatches for part stars especially at relatively faint magnitude.

\begin{table*}
\begin{center}
\caption{Comparison of the photometric metallicity estimate from SDSS $u$ and SCUSS $u$, for bright stars ($g<19$). Column (1) shows the g magnitude bins, and the rest columns in each row give the mean values of information about a bunch of stars which belong to the corresponding bins. The last two columns give the mean metallicity values estimated respectively from SDSS $u$ and SCUSS $u$.}
\begin{tabular}{|c|c|c|c|c|c|c|}
\hline
\hline
\multirow{2}*{Interval of $g$ magnitude} & \multirow{2}*{SDSS $u_{mean}$} & \multirow{2}*{SCUSS $u_{mean}$} & \multirow{2}*{$g_{mean}$} & \multirow{2}*{$r_{mean}$} & [Fe/H]$_{mean}$ & [Fe/H]$_{mean}$   \\
          &            &            &            &            & from SDSS $u$ &  from SCUSS $u$   \\ \hline
16.0 $\sim$ 16.5       &  17.63     &  17.79     &  16.44     &  15.94     &  -0.67     &  -0.85 \\ \hline
16.5 $\sim$ 17.0       &  18.10     &  18.10     &  16.76     &  16.27     &  -0.90     &  -0.93 \\ \hline
17.0 $\sim$ 17.5       &  18.55     &  18.55     &  17.26     &  16.78     &  -0.94     &  -0.94 \\ \hline
17.5 $\sim$ 18.0       &  19.04     &  19.04     &  17.75     &  17.27     &  -1.02     &  -1.01 \\ \hline
18.0 $\sim$ 18.5       &  19.56     &  19.55     &  18.26     &  17.76     &  -1.16     &  -1.16 \\ \hline
18.5 $\sim$ 19.0       &  20.05     &  20.05     &  18.76     &  18.27     &  -1.39     &  -1.35 \\ \hline
\hline
\end{tabular}
\label{Metallicity_comparison}
\end{center}
\end{table*}

\section{The overdensities in the south Galactic cap}

\par
\begin{figure*}
\begin{center}
\includegraphics[width=0.7\hsize]{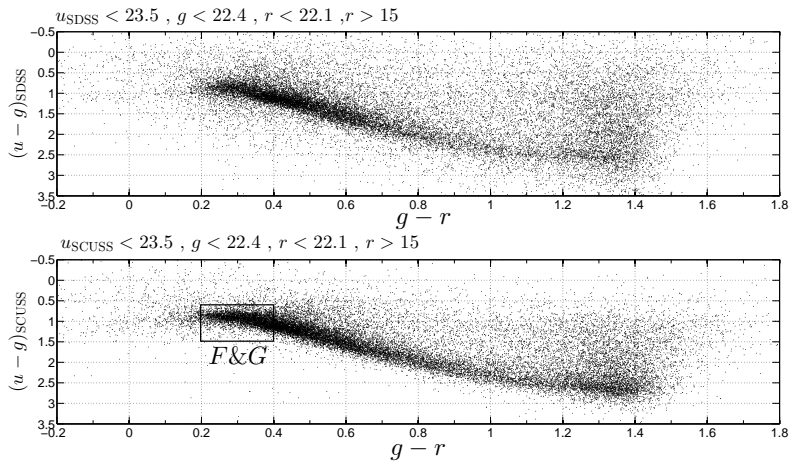}
\caption{$u-g$ vs. $g-r$ two-color diagram of 50,000 stars randomly selected from the SCUSS database. The upper figure is plotted using SDSS $u$ with selection criteria $u<23.5, g<22.4, r<22.1, r>15$, while the bottom one using SCUSS $u$ with the same criteria. Here, we choose F/G stars as tracers from the color box $u-g>0.6,~0.2<g-r<0.4$ as indicated by the rectangular box in the bottom figure.}
\label{Double_color_figure}
\end{center}
\end{figure*}

 The SCUSS covers the south Galactic cap ($30^\circ<l<210^\circ,~ -80^\circ<b<-20^\circ$) and it stretches a wide range along the celestial equator.  As shown in Fig.~\ref{Double_color_figure}, we select the F/G stars ($0.2<g-r<0.4$) as sample stars to trace the overdensities. In contrast, we plot two two-color diagrams $u-g$ vs. $g-r$, the upper figure is plotted using SDSS $u$ with selection criteria $u<23.5, g<22.4, r<22.1, r>15$, while the bottom one using SCUSS $u$ with the same criteria. We can also clearly find that the stars in the upper diagram which uses SDSS $u$ are more scattered, especially in the blue end. It provides an illustration of reduced statistical errors of SCUSS $u$. We also use the condition $u-g>0.6$ to remove the quasars. In addition, since SCUSS $u$ has the limit magnitude of $23.5$, whereas SDSS $u$ has the limit magnitude of $22.1$, we expect to pick out more sample stars from SCUSS database. We have selected F/G stars from 50,000 stars in SCUSS database and find that there are 4,566 F/G stars when using SCUSS $u$ with $u<23.5$, but only 3,572 stars when using SDSS $u$ with $u<22.1$.

\begin{figure*}
\includegraphics[width=1.0\hsize]{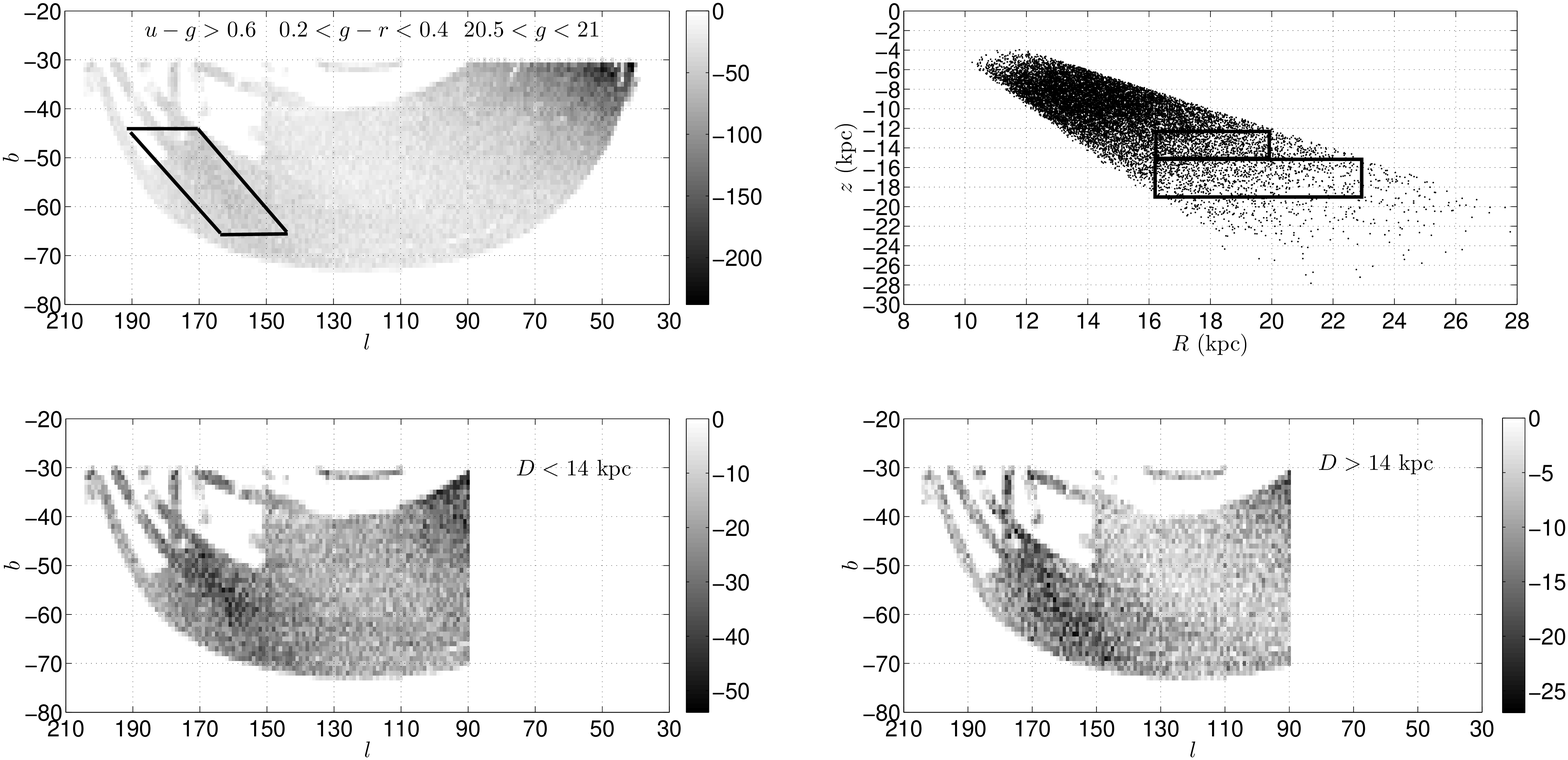}
\caption{Top left panel: Spatial distribution of the F/G stars with $20.5<g<21$. Top right panel: Distribution of the F/G main sequence stars with $20.5<g<21$ in the Sgr stream area (surrounded by parallelogram) in cylindrical galactocentric coordinate system. Bottom two panel: Spatial distribution of F/G stars with heliocentric distance $D<14$ kpc and $D>14$ kpc, respectively. The three gray-scale bars all indicate star count density in each pixel ($1^{\circ} \times 1^{\circ}$).}
\label{Overdensity}
\end{figure*}

\par After comparing with F/G stars spatially in different magnitude bin, we find that the overdensity is spatially obvious in the range of $20.5<g<21$ for the SCUSS observation, as shown in the top left panel in Fig.~\ref{Overdensity}. From Fig.~\ref{Overdensity}, we can see two overdensities: Sgr stream and Hercules-Aquila cloud which were discovered previously \citep[][]{Ibata94,Newberg02,Belokurov07}. But the Sgr stream in the top left panel of Fig.~\ref{Overdensity} is less distinct, we select the Sgr stream stars further by distance cut-off. It consists of the following procedures:

\begin{itemize}
\item[(1).] Selecting main sequence stars by rejecting those objects at distances larger than $0.15$ mag from the stellar locus which is described by following Equation \citep{Juric08}
\begin{align}
(g-r)_0 =& 1.39\{1-exp[-4.9(r-i)_{0}^3-2.45(r-i)_{0}^2 \nonumber \\
 & -1.68(r-i)_{0}-0.05]  \} \nonumber
\end{align}
\item[(2).] \textbf{Computing the heliocentric distance of each star from the following equations, the parallax relation implicitly takes metallicity effects into account.} \citep{Juric08}
\begin{align}
M_r=&4.0+11.86(r-i)_{0}-10.74(r-i)_{0}^2  \nonumber \\
  &+5.99(r-i)_{0}^3-1.20(r-i)_{0}^4  \nonumber
\end{align}
\begin{align}
D=10^{(r-M_{r})/5+1}  \nonumber
\end{align}
\item[(3).] Transforming the coordinate ($l,~b,~D$) into the right-handed Cartesian galactocentric coordinate ($x,~y,~z$) \citep{Juric08}
\begin{align}
&x=R_\circleddot-D\cos{(l)}\cos{(b}) \nonumber \\
&y=-D\sin{(l)}\cos{(b)} \nonumber \\
&z=D\sin{(b)} \nonumber
\end{align}
\item[(4).] Transforming the coordinate ($x,~y,~z$) into the cylindrical galactocentric coordinate ($R,~\phi,~z$) \citep{Bond10}
\begin{align}
&R=\sqrt{x^{2}+y^{2}}\nonumber \\
&\phi=\arctan{(y/x)} \nonumber \\
&z=D\sin{b} \nonumber
\end{align}
\end{itemize}

\par Through the above procedures, the F/G main sequence stars with $20.5<g<21$ in the Sgr stream area are picked out, which is surrounded by parallelogram in top left panel of Fig.\ref{Overdensity}.
At the same time, we give the spatial distribution of these sample stars in cylindrical coordinate system (top right panel of Fig.\ref{Overdensity}). These stars are mainly in the eighth octant with $-22.5^\circ<\phi<0^\circ$. In order to enhance the contrast of the Sgr stream, we remove the overdensity of Hercules-Aquila cloud and divide the stars into two groups by the heliocentric distance of $D=14$ kpc. Here, we also plot the spatial distribution of stars in ($l,~b$) panel (bottom two panels in Fig.~\ref{Overdensity}). We find that the Sgr stream stars are obvious when $D> 14$ kpc. In the following study of the MDF in the Sgr stream, we mainly select those sample stars in the range of $D>14$ kpc which is shown by the two rectangles in the top right panel of Fig.~\ref{Overdensity}.

\section{Photometric metallicity estimation of distant stars in south Galactic cap}

\par
\begin{figure*}
\includegraphics[width=1.0\hsize]{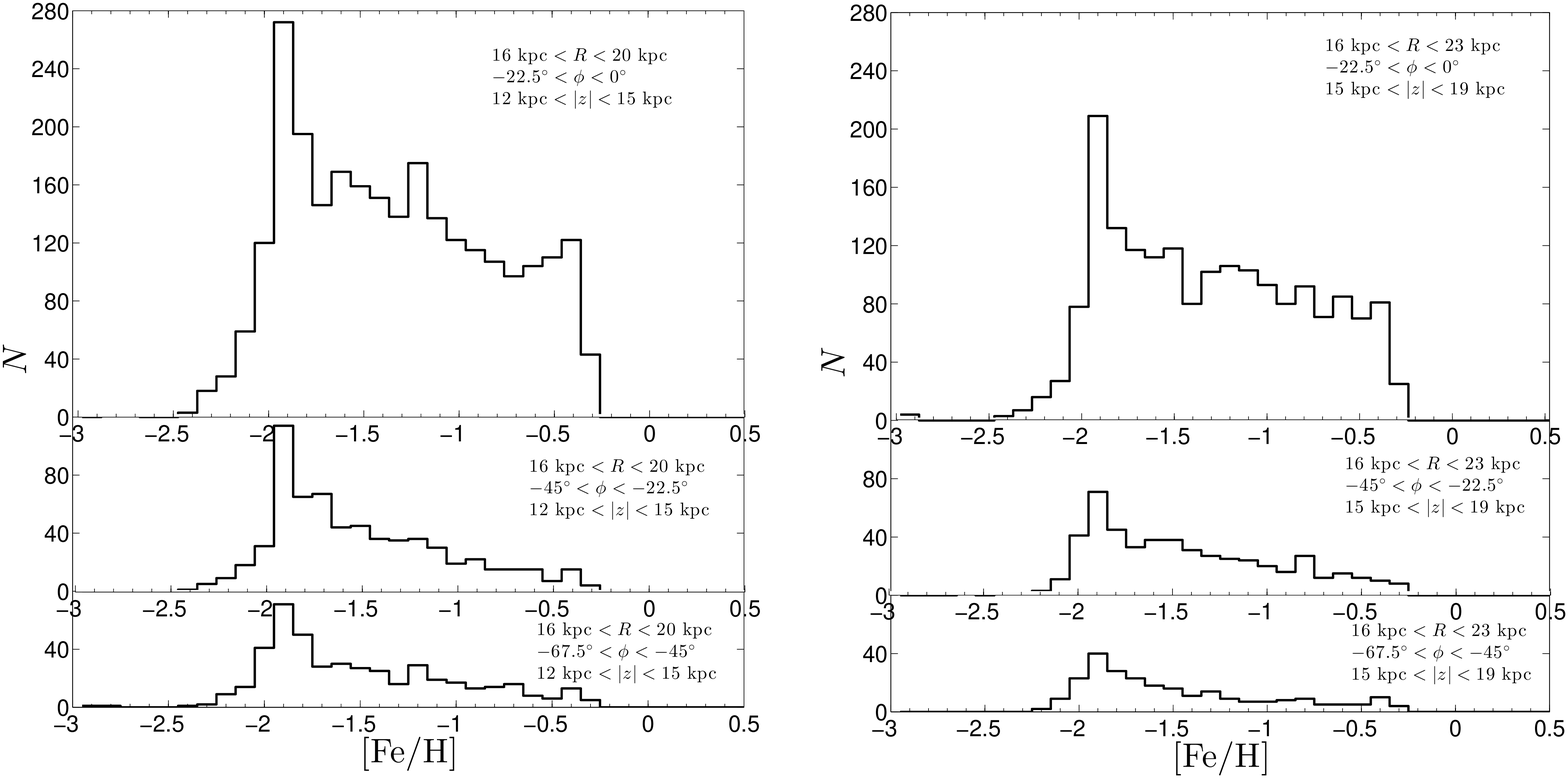}
 \caption{The metallicity distribution of the F/G ($0.2<g-r<0.4$) main sequence stars with $g<21$ in different space volume. The top two histograms correspond to the Sgr stream region. The middle and the bottom histograms correspond to the vicinities of the Sgr stream region.}
\label{Metallicity}
\end{figure*}

 Since the photometric metallicity estimate we have derived in Section 3
can also be used up to the faint magnitude of $g=21$. We may use the photometric metallicity
calibration to analyze the metallicity distribution of the sample stars which contain large number of Sgr steam stars.
As discussed above, we have selected the sample stars via cut-off on color index $g-r$ and distance.
The F/G main sequence stars with $g<21$ in the space
volume ($16$ kpc $<R<20$ kpc, $-22.5^\circ<\phi<0^\circ$, $12$ kpc $<\left| z\right| <15$ kpc)
and ($16$ kpc $<R<23$ kpc, $-22.5^\circ<\phi<0^\circ$, $15$ kpc $<\left| z\right| <19$ kpc)
are chosen to evaluate the MDF, as illustrated in the top two panels of Fig.~\ref{Metallicity}.
The top two histograms in Fig.~\ref{Metallicity} correspond to the Sgr stream region.
In order to have contrasts, the stars in other space volume with angle bin $\Delta\phi=22.5^\circ,~45^\circ$
are also selected to evaluate the metallicity distribution,
as illustrated in the middle and bottom panels of Fig.~\ref{Metallicity},
which correspond to the vicinities of the Sgr stream region.
It is clear that the Sgr stream has a wide metallicity distribution, and has much more stars from
[Fe/H]$\sim-1.5$ to [Fe/H]$\sim-0.5$.

\par
\begin{figure}
\includegraphics[width=1.0\hsize]{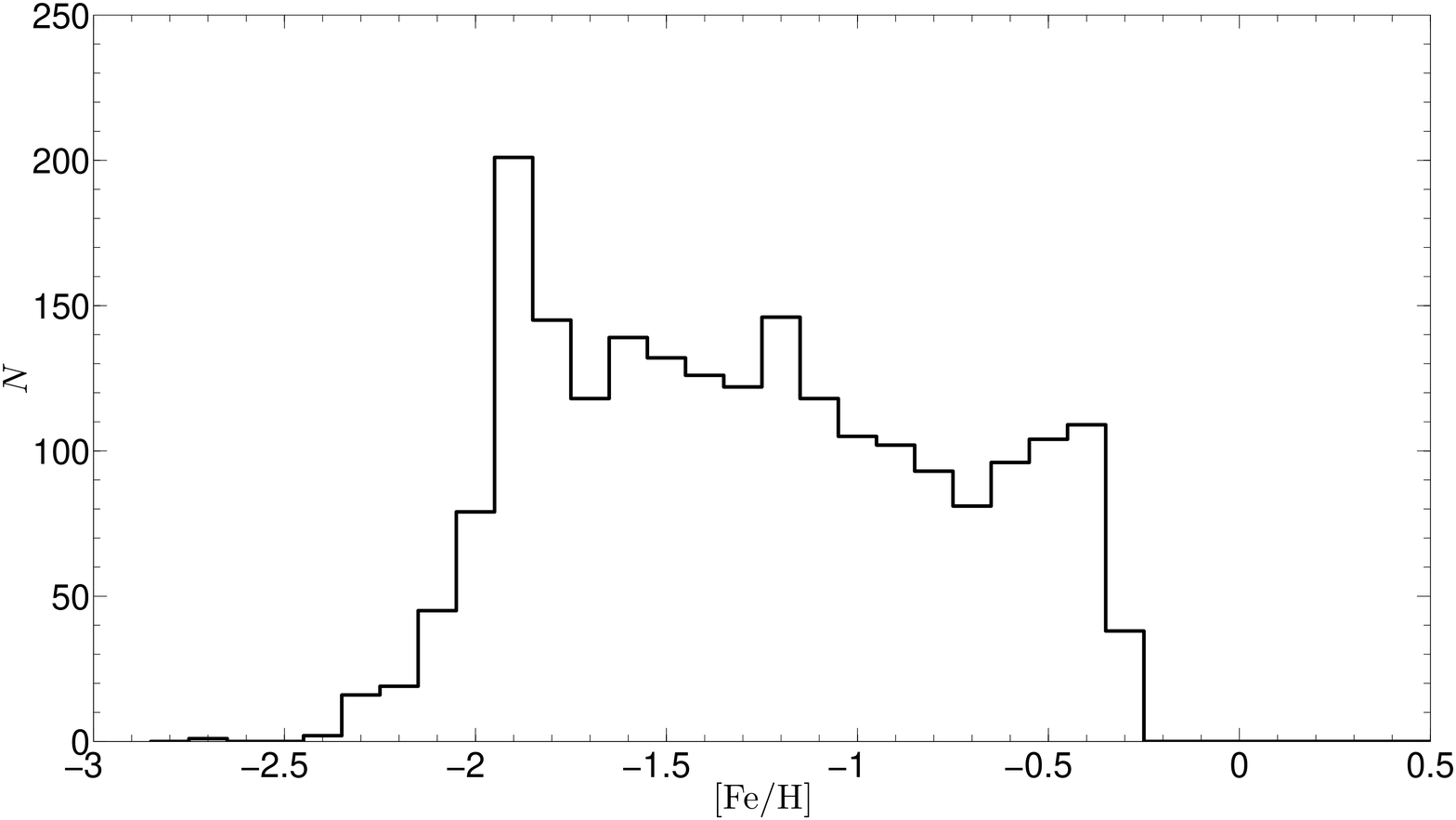}
\caption{The derived MDF by subtraction of MDF of halo stars (bottom left panel of Fig.~\ref{Metallicity}) from the Sgr region stars (top left one). It shows the wide metallicity distribution from [Fe/H]$=-0.4$ to [Fe/H]$=-2.0$.}
\label{Metallicity_sgr}
\end{figure}

\begin{figure}
\includegraphics[width=1.0\hsize]{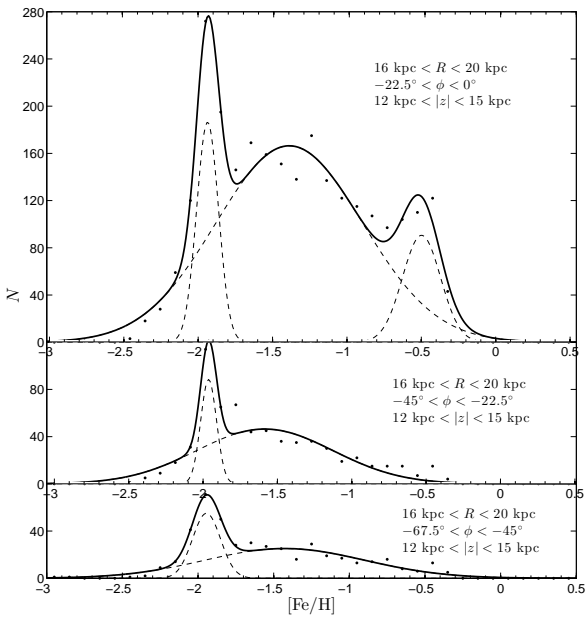}
\caption{The Gaussian models of the left half MDFs in Fig.~\ref{Metallicity}. The top MDF can be modeled by three Gaussians with peaks respectively at [Fe/H]$=-1.9$, [Fe/H]$=-1.5$ and [Fe/H]$=-0.5$. The bottom two MDFs can be modeled by two Gaussian, each with peaks at [Fe/H]$=-1.5$ and [Fe/H]$=-1.9$ respectively.}
\label{Metallicity_gauss}
\end{figure}

 Because the neighboring regions selected have the same volume as Sgr stream region,
the rough way for evaluating the MDF of the Sgr stream in the selected region is to subtract the MDF of halo stars
from the MDF of the mixed population in the equal volume space. As shown in Fig.~\ref{Metallicity_sgr},
we subtracted the MDF of halo (bottom left panel of Fig.~\ref{Metallicity})
from the mixed population (top left panel of Fig.\ref{Metallicity}).
The derived MDF can be assumed to be contributed mainly from the Sgr stream of the selected region.
We find that the Sgr stream exhibit a relatively rich metallicity distribution.
To further get the real characteristics of the MDF in the Sgr stream and halo stars,
we adopted the mixed model to fit the photometric metallicity distribution.
We fitted the left half MDFs in Fig.~\ref{Metallicity}.
We find that three-Gaussian model is appropriate for the MDF of mixed population (Sgr stream and halo stars),
and two-Gaussian model for MDF of neighbouring stars (mainly the halo stars), as shown by Fig.~\ref{Metallicity_gauss}.
The two-Gaussian of the MDF of halo stars are with peaks at [Fe/H]$=-1.5$ and [Fe/H]$=-1.9$ respectively.
This profile of the MDF of the halo stars can be explained by the argument by \cite{Carollo07}
that the halo comprises two broadly overlapping structural components, an inner and an outer halo.
The three Gaussians of the mixed population are
with peaks at [Fe/H]$=-1.9$, [Fe/H]$=-1.5$ and [Fe/H]$=-0.5$ respectively.
By contrast, we can easily find that the extra stars at [Fe/H]$=-0.5$
reasonably belongs to the Sgr stream. However, the stars around [Fe/H]$=-1.5$
may be mixed population of Sgr stream and the inner halo. Similarly,
the Gaussian with peak at [Fe/H]$=-1.9$ is contributed jointly by Sgr stream and outer halo.
The parameters of the Gaussians for fitting are shown in Table~\ref{Parameter}.

\begin{table}
\begin{center}
\caption{The parameters of the Gaussians for fitting the MDFs.}
\begin{tabular}{p{2.2cm}p{2.2cm}p{2.2cm}}
\hline
\hline
\multicolumn{3}{c}{$N=a1\cdot Exp{[-(\frac{x-b1}{c1})^{2}]}+a2\cdot Exp{[-(\frac{x-b2}{c2})^{2}]}+a3\cdot Exp{[-(\frac{x-b3}{c3})^{2}]}$}  \\ \hline
$a1=186.4$  & $b1=-1.937$   &  $c1=0.1022$ \\ \hline
$a2=166.4$  & $b2=-1.397$   &  $c2=0.6862$ \\ \hline
$a3=90.58$  & $b3=-0.5229$   &  $c3=0.1777$ \\ \hline
\multicolumn{3}{c}{$N=a1\cdot Exp{[-(\frac{x-b1}{c1})^{2}]}+a2\cdot Exp{[-(\frac{x-b2}{c2})^{2}]}$}  \\ \hline
$a1=88.4$  & $b1=-1.93$   &  $c1=0.0741$   \\ \hline
$a2=46.41$  & $b2=-1.559$   &  $c2=0.6376$ \\ \hline
\multicolumn{3}{c}{$N=a1\cdot Exp{[-(\frac{x-b1}{c1})^{2}]}+a2\cdot Exp{[-(\frac{x-b2}{c2})^{2}]}$}  \\ \hline
$a1=55.07$  & $b1=-1.944$   &  $c1=0.1308$   \\ \hline
$a2=25.15$  & $b2=-1.427$   &  $c2=0.7478$ \\ \hline
\hline
\end{tabular}
\label{Parameter}
\end{center}
\end{table}

\section{Summary}

\par In this paper, based on SCUSS $u$ and SDSS $g,r,i,z$,
we derive a photometric metallicity calibration for estimating
the stellar metallicity [Fe/H]$_{phot}$ by using 32,542 F- and G-type
main sequence stars from SDSS spectra. The rms scatter of
the photometric metallicity residuals relative to
spectrum-based metallicity is $0.14$ dex when $g-r<0.4$ and $0.16$ dex when $g-r>0.4$.
The rms statistical scatter of metallicity due to u-band photometric uncertainty increases
with $g$ magnitude from $0.014$ dex for $g<17$ to $0.070$ dex at $g=18.5$, $0.379$ dex
at $g=20.5$, $0.445$ dex at $g=21$ and $0.750$ dex at $g=21.3$.
So we limit the application of photometric estimates in the range of $g<21$.
This application range of [Fe/H]$_{phot}$ is wide, and it allows metallicity
to be determined for all SCUSS main-sequence stars in these criteria range.
So we can derive the [Fe/H]$_{phot}$ estimates for numerous farther stars.

\par As an example, we select Sgr stream and its neighboring field halo stars
in south Galactic cap to study its metallicity distribution.
We find that the Sgr stream around the cylindrical Galactocentric coordinate
of $R\sim 19$ kpc, $\left| z\right| \sim 14$ kpc exhibit a relative rich metallicity distribution,
and the MDF of the neighboring field halo stars in our studied fields can be modeled by two-Gaussian model, with peaks respectively at [Fe/H]$=-1.9$ and [Fe/H]$=-1.5$.

\section*{ACKNOWLEDGMENTS}
\par We especially thank the anonymous referee for his/her helpful
comments and suggestions that have significantly improved the paper.
We thank Martin C. Smith for useful discussions to improve the study
at the beginning of this work. This work was supported by joint fund
of Astronomy of the the National Nature Science Foundation of China
and the Chinese Academy of Science, under Grants U1231113. This work
was also supported by the National Natural Foundation of China
(NSFC, No.11373033, No.11373003, No.11373035, No.11203034,
No.11203031, No.11303038, No.11303043), and by the National Basic
Research Program of China (973 Program) (No. 2014CB845702,
No.2014CB845704, No.2013CB834902).

\par We would like to thank all those who participated in observations and
data reduction of SCUSS for their hard work and kind cooperation.
The SCUSS is funded by the Main Direction Program of Knowledge Innovation of
Chinese Academy of Sciences (No. KJCX2-EW-T06). It is also an international
cooperative project between National Astronomical Observatories,
Chinese Academy of Sciences and Steward Observatory, University of Arizona, USA.
Technical supports and observational assistances of the Bok telescope are
provided by Steward Observatory. The project is managed by the
National Astronomical Observatory of China and Shanghai Astronomical Observatory.

\par Funding for SDSS-III has been provided by the Alfred P. Sloan Foundation,
the Participating Institutions, the National Science Foundation,
and the U.S. Department of Energy Office of Science. The SDSS-III web site is \emph{http://www.sdss3.org/}.

\par SDSS-III is managed by the Astrophysical Research Consortium for the
Participating Institutions of the SDSS-III Collaboration including the
University of Arizona, the Brazilian Participation Group, Brookhaven
National Laboratory, Carnegie Mellon University, University of Florida,
the French Participation Group, the German Participation Group, Harvard University,
the Instituto de Astrofisica de Canarias, the Michigan State/Notre Dame/JINA Participation Group,
Johns Hopkins University, Lawrence Berkeley National Laboratory, Max Planck Institute for Astrophysics,
Max Planck Institute for Extraterrestrial Physics, New Mexico State University, New York University,
Ohio State University, Pennsylvania State University, University of Portsmouth, Princeton University,
the Spanish Participation Group, University of Tokyo, University of Utah, Vanderbilt University,
University of Virginia, University of Washington, and Yale University.

\end{document}